# DFT investigation of novel cubic carbon allotrope, “yne-$C_{16}$”.

Samir F. Matar

Lebanese German University (LGU), Sahel Alma, Jounieh, Lebanon

https://orcid.org/0000-0001-5419-358X

*email: s.matar@lgu.edu.lb

***Abstract***

*Novel cubic allotrope yne-$C_{16}$ with $sp^3/sp^1$ carbon hybridizations is devised based on crystal chemistry and computations of the ground structure and the energy-dependent quantities within the quantum density functional theory DFT. With respect to **srs**-$C_8$ characterized by trigonal $C(sp^2)$, yne-$C_{16}$ is identified with original C≡C triple bond-like units and distorted tetrahedra. The new allotrope is qualified with unknown topology. At the elementary building unit, $C_{16}$ looks like a pyramidal molecule such as **:**$PCl_3$, with “**:**“ symbolizing the phosphorous lone pair. With this conceptualization $C_{16}$ has three bond pairs (SIGMA-like) making the pyramid base and the C(triple-bond)C with large PI character resembling phosphorous electron lone pair (**:**) at the apex of the pyramid. The new allotrope was found cohesive and stable both mechanically (positive elastic constants) as well as dynamically (phonons band structure). $C_{16}$ is characterized with a weakly metallic character.*

## Introduction

The exploration of carbon allotropes with different dimensionalities (D), mainly 3D and 2D is an active field within the scientific community. The three-dimensional 3D arrangements of corner sharing *C4* tetrahedra with purely C(sp$^3$) are at the origin of the extreme hardness of diamond in the prevailing cubic form as well as in the less common hexagonal allotrope 'Lonsdaleite'. Topology-wise using TopCryst. program [1], the cubic and hexagonal allotropes are the aristo types labeled **dia** and **lon**, respectively; both characterized by large electronic band gaps. Other families of carbon allotropes follow these nomenclatures; they are documented in the SACADA data base for carbon allotropes [2]. Departing from purely C(sp$^3$), several original allotropes were identified thanks to rationalized insertions of extra carbon bringing changes into the C(sp$^3$) lattice leading to hybrid allotropes with C(sp$^2$) and/or C(sp$^1$) besides C(sp$^3$). Changes of the physical properties are consequently induced such as metallic behavior [3]. Carbon with sp$^2$ hybridization is known for paving the planes in 2D graphite. Going beyond this dimensionality, Hoffmann et al. [4] proposed 3D paving with C(sp$^2$) in body center tetragonal $C_8$ with **ths** topology, listed under N°15 in the SACADA database [5].

Nowadays several new allotropes are found using structure prediction programs like CALYPSO [6]. However, novel structures can be obtained from crystal engineering rationale as presented herein. In both research cases, the predictions of stoichiometries need validation through quantitative studies of the energies and the derived physical properties thanks to first principles calculations to identify ground state structures with all the pertaining physical properties. The quantum mechanics density functional theory (DFT) [7, 8] has been proven as a reliable framework thanks to the different approximation schemes of the exchange-correlation effects XC (*vide infra*) built throughout the years. In the present paper we follow this two-steps protocol to identify novel cubic $C_{16}$ with mixed hybridizations sp$^3$/sp$^1$ characterized by an unknown topology. Analyses of mechanical and dynamic properties revealed a stable highly compressible allotrope and the electronic band structure pointed out to a weakly metallic behavior.

## 1- Computational methodology

To identify the ground state structures corresponding to the energy minima and the subsequent prediction of the mechanical and dynamical properties we used the DFT-based Vienna Ab initio Simulation Package (VASP) code [9,10] and the projector augmented wave (PAW) method [10, 11] for the atomic potentials. DFT exchange- correlation (XC) effects were considered using the generalized gradient functional approximation (GGA) [12]. Relaxation of the atoms onto the ground state structures was performed with the conjugate gradient algorithm according to Press *et al*. [13]. The Bloechl tetrahedron method [14] with corrections according to the Methfessel and Paxton scheme [15] was used for geometry optimization and energy calculations, respectively. Brillouin-zone (BZ) integrals were approximated by a special **k**-point sampling according to Monkhorst and Pack [16]. Structural parameters were optimized until atomic forces were below 0.02 eV/Å and all stress components were < 0.003 eV/Å$^3$. The calculations were converged at an energy cutoff of 400 eV for the plane-wave basis set in terms of the **k**-point integration in the reciprocal space from $k_x(6)\times k_y(6)\times k_z(6)$ up to $k_x(12)\times k_y(12)\times k_z(12)$ to obtain a final convergence and relaxation to zero strains for the original stoichiometries presented in this work. In the post-processing of the ground state electronic structures, the charge density projections were operated on the lattice sites.

The mechanical stability criteria were obtained from the calculations of the elastic constants. The Voigt's method of averaging the elastic constants was used herein for obtaining the bulk and shear moduli [17]. For the calculation of the Vickers hardness a semi-empirical model was used [18]. The dynamic stabilities were confirmed from the phonon positive magnitudes. The corresponding phonon band structures were obtained from a high resolution of the cubic Brillouin Zone (BZ) according to Togo *et al*. [19]. The electronic band structures were obtained using the all-electron DFT-based ASW method [20] and the GGA XC functional [12]. The VESTA (Visualization for Electronic and Structural Analysis) program [21] was used to visualize the crystal structures and charge densities.

## 2- Crystal chemistry and charge densities results

### *a- Crystal Chemistry Engineering*

Alike **ths** $C_8$ [4], C($sp^2$) hybridization was also observed in cubic $C_8$ possessing the **srs** topology [5]. The polyhedral representation of the structure shown in Fig. 1a exhibits trigonal carbons. The crystal data given in Table 1 show the feature of C(8c) x,x,x Wyckoff position. Specifically, we found that the C($sp^2$) based structure with x = 1/8 (0.125) is maintained as such, up to x = 0.23, i.e. just below ¼, then it becomes purely **dia** C($sp^3$). In the other direction, the structure remains trigonal down to x = 0.07 then it becomes **dia** C($sp^3$). Therefore, with x = 0, ¼ or ½ one gets diamond structure (**dia**) with only C($sp^3$). Based on these observations novel cubic $C_{16}$ was derived with two C(8c) positions. The two **srs** $C_8$ and $C_{16}$ stoichiometries and novel $C_{16}$ were submitted to full geometry optimizations to ground state structure and energy. The resulting structures are shown in Figure 1 and the crystal parameters are given in Table 1.

Reporting firstly on $C_8$, characterized in body center cubic (bcc) system with space group $I4_132$ No.214, a good agreement with the published value of lattice constant can be observed (cf. SACADA No. 5). The converged lattice parameters of novel $C_{16}$ characterized in lower symmetry simple cubic (sc) $P4_132$ No.213 show a larger lattice constant and subsequent larger volume which remains nevertheless less than twice the volume of $C_8$. This is translated by a larger density of $C_{16}$ with $\rho(C_{16})$ = 2.47 versus $\rho(C_8)$= 2.29 g/cm$^3$. The cohesive energy obtained by subtracting the atomic energy of C in a large box (-6.6 eV) from the total energy of the allotrope shows a more cohesive $C_8$ than $C_{16}$. Yet, both can be considered as metastable versus graphite and diamond. Regarding the x values of C1 and C2 at (8c) position, they are both larger than in $C_8$ while remaining between 1/8 and ½; i.e. 1/8 < x(C1)= 0.147 < ¼ < x(C2)= 0.307 < ½. An unknown topology was assigned to $C_{16}$ from TopCryst. online analysis [5] letting propose it as a new carbon allotrope.

The geometry optimized structures in Figure 1 are sketched in ball-and-stick and polyhedral projections for more visibility. Also, a projection along the "*A3*" cube large diagonal is provided. Although the validity of the structure is not questioned, the feature of discontinuous atomic arrangement is highlighted in **srs** $C_8$, oppositely to $C_{16}$. Despite the established cohesiveness, such a feature could generate instability in the physical properties as shown in next section regarding the mechanical as well as the dynamic properties. In $C_{16}$ two carbon sites are shown with two distinct atomic colors. The brown color spheres (C1) are like those of pristine $C_8$ that become connected thanks to the white spheres (C2). Chemistry-wise the linear arrangement C-C-C-C is relevant to C-C≡C-C, i.e., with acetylene-like triple bond C≡C. Indeed, d(C2-C2) = 1.20 Å is a magnitude found in linear acetylene molecule $C_2H_2$. We also observed such short distances in yne-$C_8$ [3]. The tetrahedra are characterized by a larger distance d(C1-C2) = 1.40 Å.

Looking more closely, the tetrahedron in $C_{16}$ is strongly distorted with a large departure from the ideal ∠C-C-C angle of 109.47°. Lower and higher values are observed: 94° and 119°. These magnitudes close to right angle (90°) and trigonal angle (120°) pertain to trigonal pyramid shape observed in molecules such as **:**$PCl_3$ with phosphorous at the apex and three Cl atoms forming at the corners of a trigonal base. Upon completing the geometry with phosphorous lone pair LP (**:**), the structure resembles a distorted tetrahedron. With respect to schematic description of **:**$PCl_3$ above, $C_{16}$ representation is completed with C2≡C2 (white spheres) at the apex at the of the pyramid as shown at the right-hand side of Fig. 1b within a projection along the cube *A3* diagonal (vertical direction). The description will be further illustrated with a qualitative description of the charge density projections (*vide infra*).

*b- Charge density projection*

At this point it becomes relevant to establish a qualitative chemical rationale regarding the distribution of the charges in the two allotropes under investigation, particularly the new one $C_{16}$. Figure 2 shows the respective charge density projections with yellow volumes. At the crossing of a plane red spots and green spots indicate large and low charge concentrations respectively. In $C_8$ (Fig. 2a) there is a continuous charge density with yellow volumes -except for the separated trigonal motif shown in Fig. 1a right-hand representation along the cube A3 diagonal- around and between the atoms that can be assigned to the $sp^2$ hybridization throughout the structure. In $C_{16}$ (Fig. 2b) differences of charge distributions are observed with large yellow envelopes around the C2-C2 pairs assigned to C≡C characterized by C($sp^1$) i.e., with one σ bond and two π bonds. Such concentrated/localized large charge around C≡C leaves little charge density for the remaining end carbon in the linear C1-C2≡C2-C1 subunit, such C1 being part of the tetrahedron constitution. Then it is expected that at crossing the cell face one observed green spots, oppositely to the red spots shown in $C_8$ projection where the charge is more homogeneously distributed.

Connecting with the discussion at the end of above subsection related to relationship of $C_{16}$ with $:PCl_3$ pyramidal molecule, the charge density projections reproduce to some extent the chemical picture. This is especially visualized in the projections along the *A3* diagonal of the cube shown in Fig. 2b last two representations with ball-and-stick and polyhedral projections: There are three bond pairs ($\sigma$-like) C1-C1 making the pyramid base with $\angle$~120° and the C2≡C2 with large charge density due to prevailing $\pi$ character on top of the pyramid representing the high charge density LP (**:**) present in $:PCl_3$, whence the asymmetric charge distribution.

3- **Mechanical and dynamic stabilities**

The mechanical stability was inferred from the calculation of the elastic tensors by inducing finite distortions of the lattice. In the cubic system there are three distinct elastic constants $C_{11}$, $C_{12}$ and $C_{44}$ (expressed in units of GPa) which must be all positive with the condition $C_{11} > C_{12}$.

For srs-$C_8$: $C_{11}$= 218, $C_{12}$= 282, and $C_{44}$= - 289, checked with different calculations. The fact that $C_{11} < C_{12}$ and the negative value of $C_{44}$ are two indications of mechanically unstable system. Further analysis of $C_8$ was discarded.

For $C_{16}$: $C_{11}$= 162, $C_{12}$= 58, and $C_{44}$ = 65, the conditions of mechanical stability are met, letting calculate the bulk B and shear G moduli:

$$B = 1/3\ (C_{11}+2C_{12}) = 93\ \text{GPa};$$

$$G = 1/5\ (C_{11}-C_{12}+3C_{44}) = 60\ \text{GPa}.$$

The Vickers hardness ($H_V$) calculated using Tian et al. [18] approach:

$$H_V = 0.92\ (G/V)^{1.137}\ G^{0.708}$$

that includes the Pugh ratio G/V amounting to 0.64 pointing to a ductile material, provides a low $H_V$= 10 GPa of a very soft system. Such behavior could be related to the molecular related character discussed above.

Regarding the dynamic stability, a relevant criterion is inferred from the properties of phonons defined as quanta of vibrations with a quantized energy thanks to the Planck constant 'h' used in its reduced form, i.e., $\hbar$ with $\hbar = h/2\pi$, resulting into by $E = \hbar\omega$ where $\omega$ is the frequency. In Fig. 3 the phonons bands are represented with red lines developing along the major directions of the simple cubic Brillouin Zone BZ (horizontal *x*-axis) separated by vertical lines for clarity. The frequency along the y-axis is expressed in units of Terahertz (THz, with 1 THz = 33cm$^{-1}$). There are 3N-3 optical modes found at higher energy than three acoustic modes that start from zero energy ($\omega = 0$) at the $\Gamma$ point (BZ center), up to a few Terahertz. Such low frequency modes correspond to the lattice rigid translation modes of the crystal (two transverse and one longitudinal which were found largely negative in **srs** $C_8$ as shown in Fig. 3a, oppositely to Fig. 3b. This result is another verification of its instability.

Focusing on novel $C_{16}$ and starting from the $\Gamma$ point, three acoustic bands can be observed up to 8 THz, and then little dispersed bands develop up to 20 THz assigned to the low frequency optic modes for vibration within the pyramidal distorted tetrahedra. At higher frequencies flat bands are observed around 30, 40, and 70 THz, corresponding to the short interatomic distances especially for the linear C1-C2≡C2-C1 where d(C2≡C2) =1.20 Å alike in acetylene $C_2H_2$ has the highest frequency line at 70 THz, then d(C1-C2) = 1.40 Å relevant to C(sp$^3$) at 40 THz as observed generally in C(sp3) systems.

**4- Electronic band structure**

The electronic band structure of $C_{16}$ was calculated using the all-electrons DFT-based augmented spherical method (ASW) [20] using the crystal structure data in Table 1. The results are shown in Figure 4. The bands (blue lines) develop along the main directions of the cubic Brillouin zone as for the phonon bands (Fig. 3). The zero energy along the vertical axis is considered with respect to the Fermi level $E_F$ with band crossing the Fermi level with a trend to a weakly metallic behavior.

## Conclusion

Based on body center cubic **srs**-$C_8$ characterized by trigonal C(sp$^2$) in 3D packing, novel $C_{16}$ was devised in simple cubic symmetry and two distinct carbon sites: C1 building pyramidal units and C2 exhibiting high charge density C2≡C2. Furthering on the peculiar features of the new allotrope found with unknown topology, a qualitative correspondence is proposed with pyramidal **:**$PCl_3$ molecule with three bond pairs ($\sigma$-like) making the pyramid base with C1 and the C2≡C2 with large $\pi$ character resembling phosphorous electron lone pair (**:**) on top of the pyramid. The new allotrope with mixed $sp^3/sp^1$ carbon hybridization was found cohesive and stable mechanically (positive elastic constants) as well as dynamically (phonons band structure). A weakly metallic behavior was identified from electronic band structure analysis.

*Note: The results obtained herein regarding **srs** $C_8$ invalidating its mechanical and dynamic stabilities could not be confronted with literature that we searched extensively. However, the more complex $C_{16}$ having been studied and found stable in the same conditions of high precision energy and energy dependent calculations provide confidence to our findings*.

**References.**

Table 1. Crystal structure parameters cubic carbon allotropes: **srs** $C_8$ and present work (pw) $C_{16}$

| Topology<br><br>Space Group | $C_8$ **srs** topology SACADA#5<br>*I*$4_1$32; No.214 | *pw* $C_{16}$<br>unknown topology<br>*P*$4_1$32. No.213 |
|---|---|---|
| *a*, Å<br>c, Å | 4.115 (4.114) | 5.056 |
| $V_{cell}$, $Å^3$ | 69.66 | 129.25 |
| *Shortest d( C-C)* Å | 1.45 | 1.20; 1.40; 1.80 |
| Atomic position | C1(8c) 0.125, x, x | C1(8c) 0.147, x, x<br>C2(8c) 0.307, x, x |
| Density g/cm$^3$<br><br>E(coh.)/atom eV | 2.29<br><br>-1.28 | 2.47<br><br>-1.03 |

FIGURES

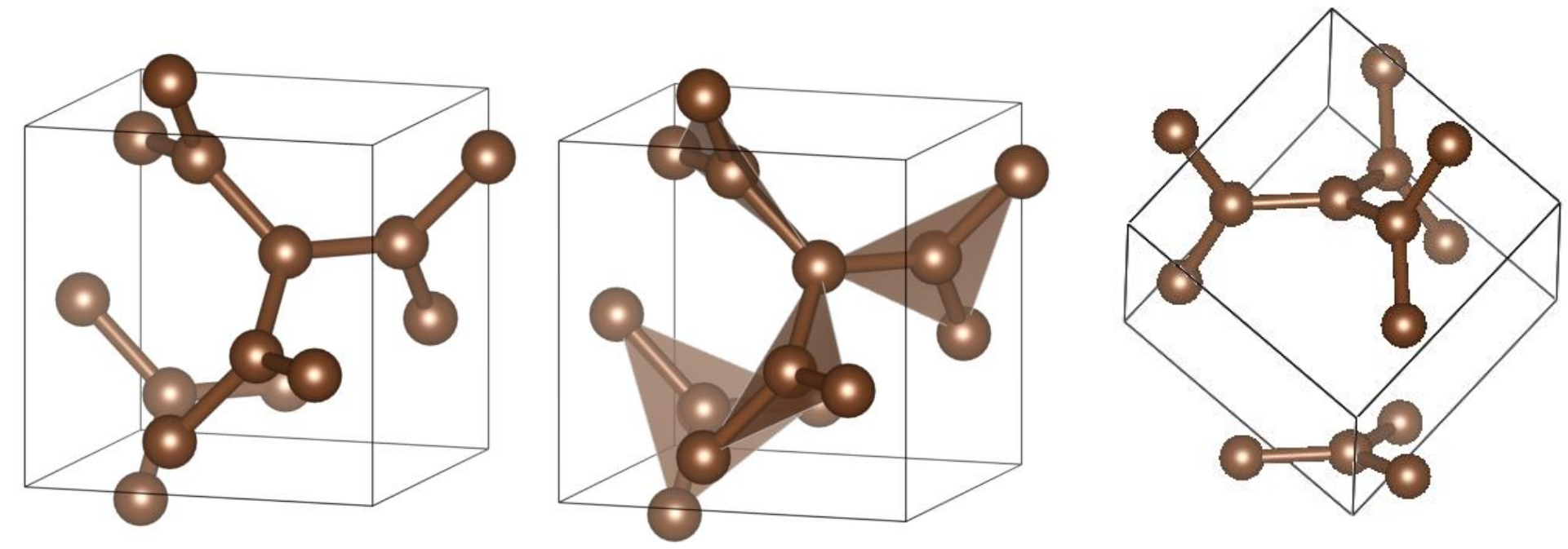

a) **srs** $C_8$

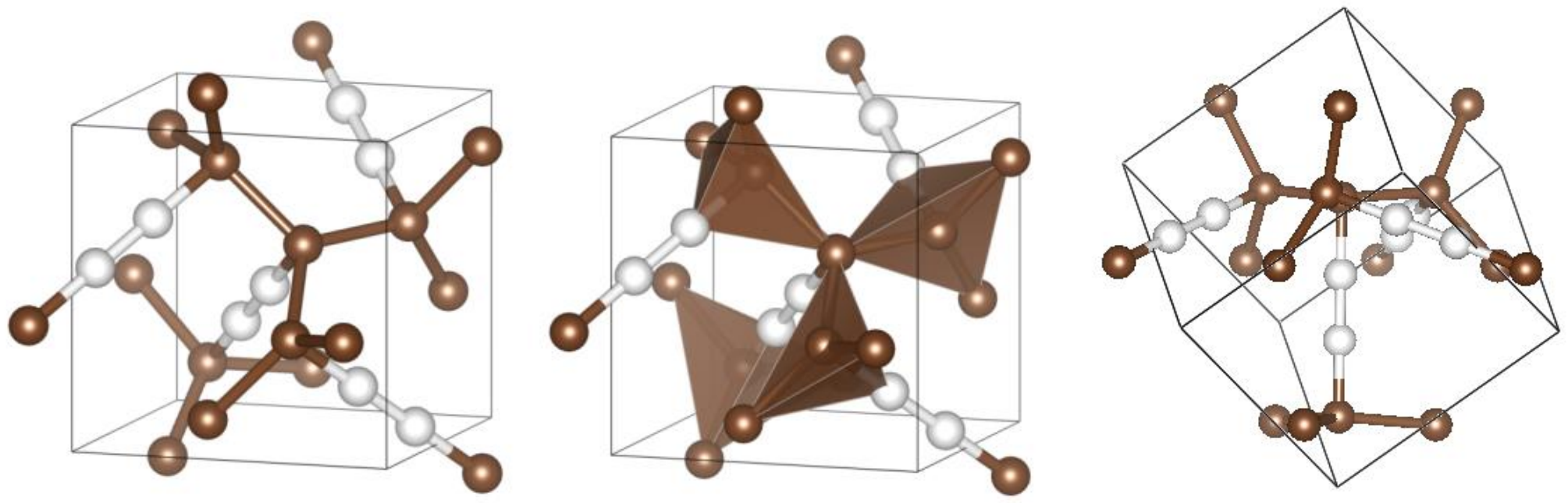

b) $C_{16}$

Figure 1. Sketches of the crystal structures in ball-and-stick and polyhedral projections for a) **srs** $C_8$, b) $C_{16}$ (brown and white spheres correspond to pyramid-tetrahedral $sp^3$ C1 and $sp^1$ C2 respectively in Table 1). The right-hand side projections along the *A3* cube diagonal (right) show the connectivity.

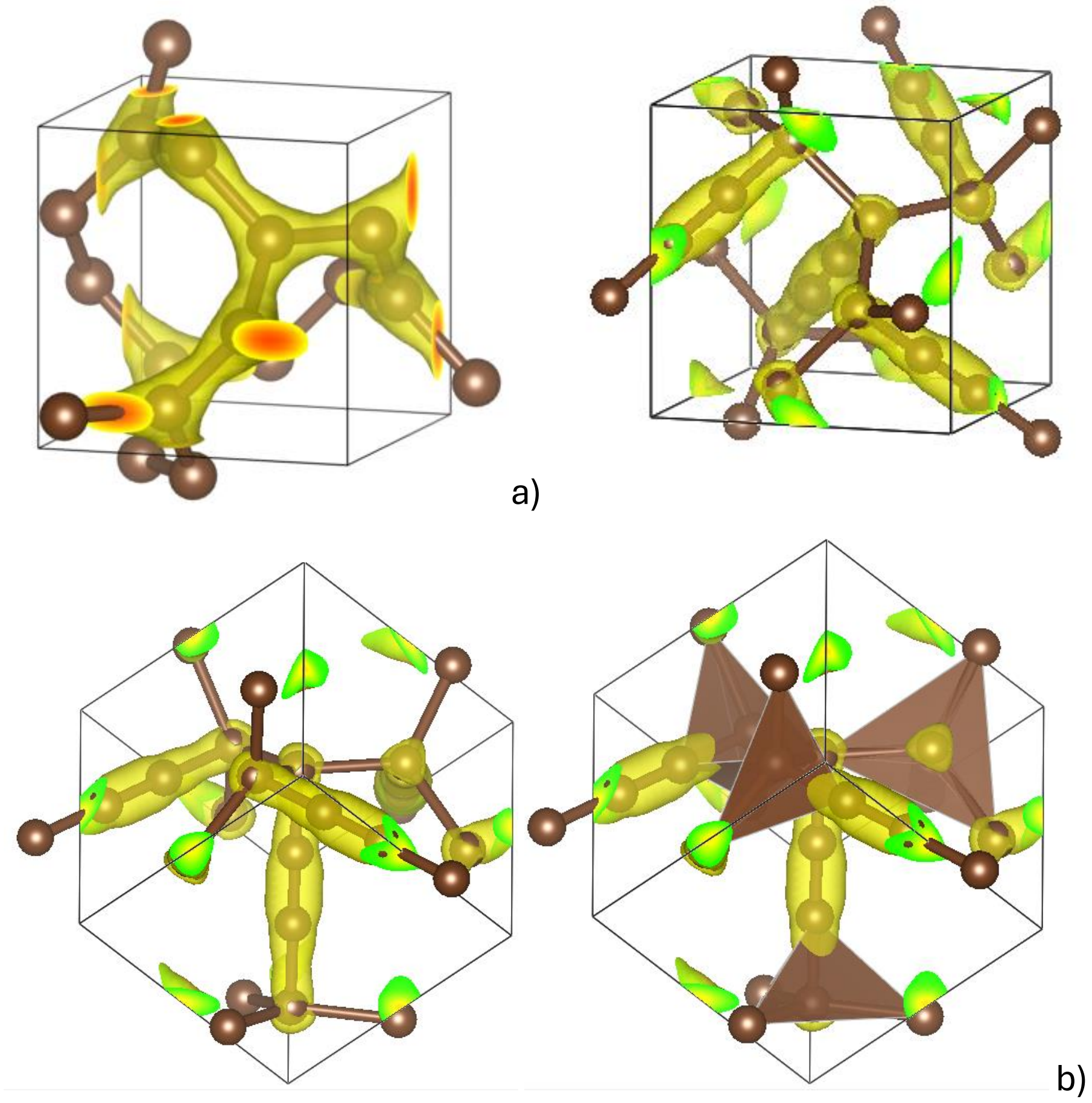

Figure 2. Charge density in a) **srs** c-$C_8$, b) $C_{16}$ with orthogonal projection and two projections along the *A3* cube diagonal for better visualization of the C≡C high density.

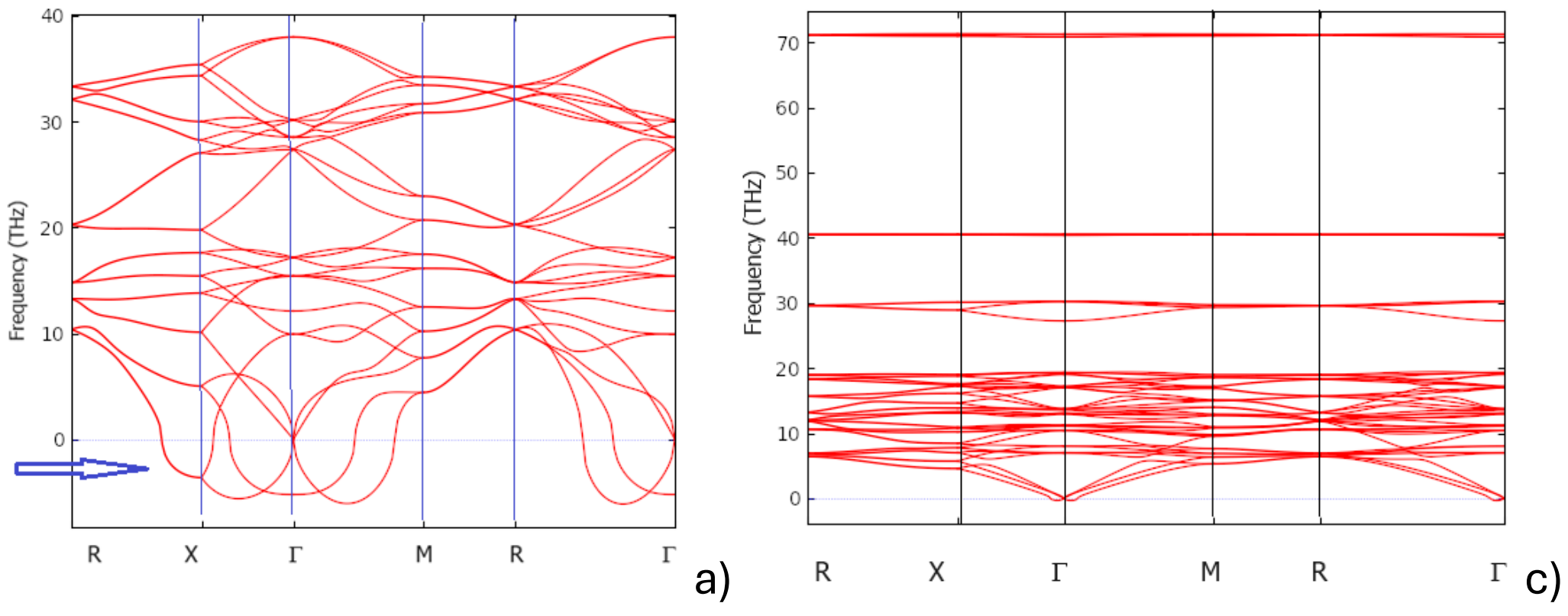

Figure 3. Phonons band structures of a) $C_8$ (the blue arrow points to largely negative acoustic phonon mode =dynamically unstable system), and b) $C_{16}$.

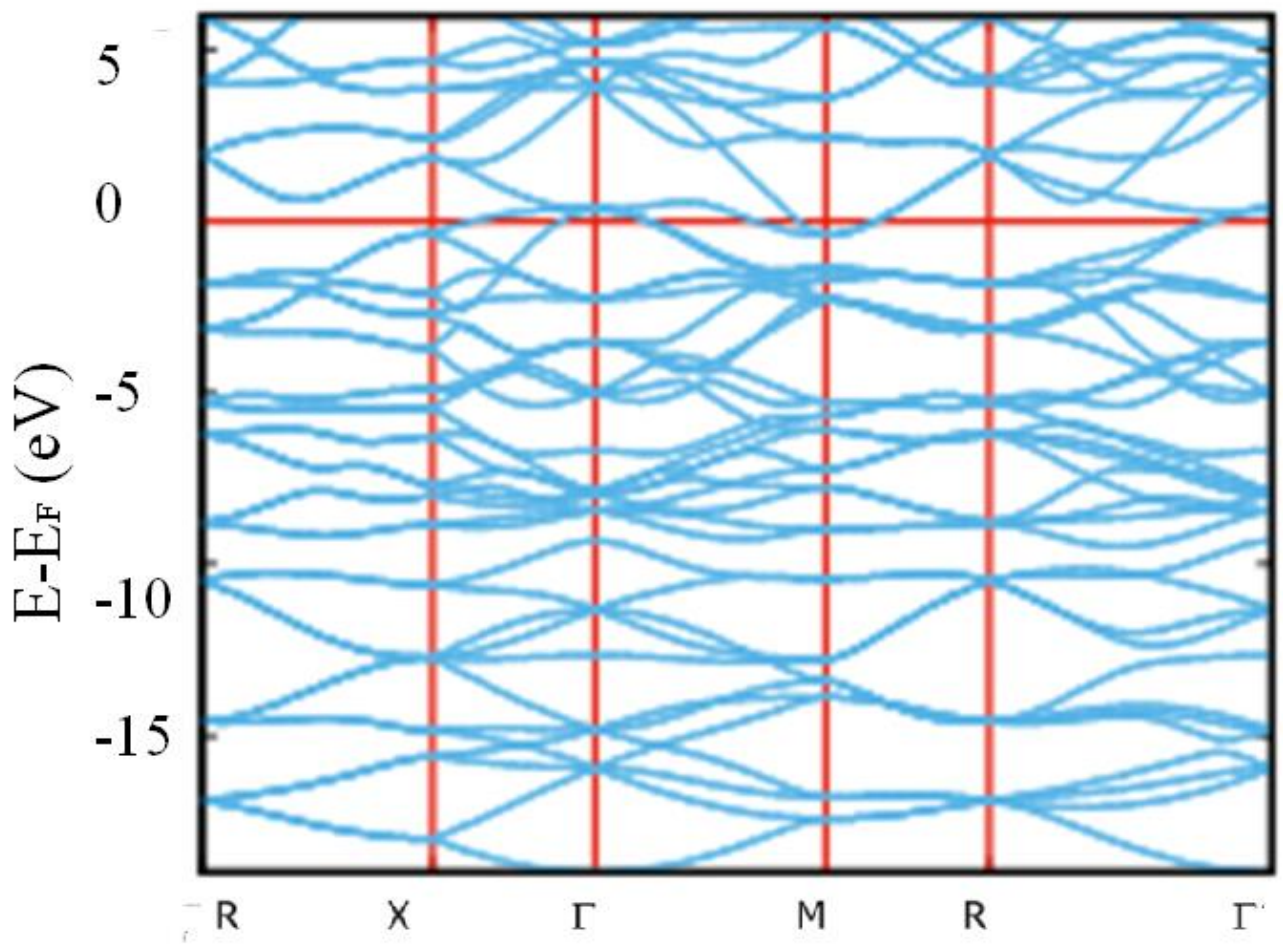

Figure 4. Electronic band structure of $C_{16}$